\pdfoutput=1

\documentclass[11pt,twoside,a4paper,cmspaper,final,collab]{cms-tdr}
\def\svnVersion{160066}\def\cmsCernNo{2012-116}\def\cmsCernDate{\today}\def\cmsMessage{Submitted to Physical Review Letters}

\begin{document}\cmsNoteHeader{HIN-11-003}

\hyphenation{had-ron-i-za-tion}
\hyphenation{cal-or-i-me-ter}
\hyphenation{de-vices}

\RCS$Revision: 160066 $
\RCS$HeadURL: svn+ssh://alverson@svn.cern.ch/reps/tdr2/papers/HIN-11-003/trunk/HIN-11-003.tex $
\RCS$Id: HIN-11-003.tex 160066 2012-12-04 18:24:04Z alverson $
\newlength\cmsFigWidth
\ifthenelse{\boolean{cms@external}}{\setlength\cmsFigWidth{0.95\columnwidth}}{\setlength\cmsFigWidth{0.6\textwidth}}
\cmsNoteHeader{HIN-11-003} 
\title{Measurement of the pseudorapidity and centrality dependence of the transverse energy density in PbPb collisions at \texorpdfstring{$\sqrt{s_{_\mathrm{ NN}}}=2.76\TeV$}{sqrt(s[NN])= 2.76 TeV}}

\date{\today}

\abstract{
The transverse energy (\ET) in PbPb collisions at 2.76\TeV nucleon-nucleon center-of-mass energy ($\sqrt{s_{_\mathrm{NN}}}$) has been measured over a broad range of pseudorapidity ($\eta$) and collision centrality using the CMS detector at the LHC.  The transverse energy density per unit pseudorapidity ($\rd\ET/\rd\eta$) increases faster with collision energy than the charged particle multiplicity. This implies that the mean energy per particle is increasing with collision energy. At all pseudorapidities the transverse energy per participating nucleon increases with the centrality of the collision. The ratio of transverse energy per unit pseudorapidity in peripheral to central collisions varies significantly as the pseudorapidity increases from $\eta = 0$ to  $|\eta| = 5.0$. For the 5\% most central collisions the energy density per unit volume is estimated to be about $14\GeV\!/\mathrm{fm}^3$  at a time of 1\unit{fm}$\!/\!c$ after the collision. This is about 100 times larger than normal nuclear matter density and a factor of 2.6 times higher than the energy density reported at $\sqrt{s_{_\mathrm{NN}}} = 200$\GeV at RHIC.
}

\hypersetup{%
pdfauthor={CMS Collaboration},%
pdftitle={Measurement of the pseudorapidity and centrality dependence of the transverse energy density in PbPb collisions at sqrt(s[NN])= 2.76 TeV},%
pdfsubject={CMS},%
pdfkeywords={CMS, physics, heavy ion, energy distribution}}

\maketitle 

 The goal of relativistic heavy-ion collisions is to study the behavior of quarks and gluons under extreme conditions  of pressure, density, and temperature, such as those that existed shortly after the Big Bang. Similar conditions can be reproduced in the laboratory by colliding heavy nuclei at the highest possible energies. Experiments at the Relativistic Heavy Ion Collider (RHIC) have shown that at nucleon-nucleon center-of-mass energy of $\sqrt{s_{_\mathrm{NN}}} = 200$\GeV a strongly interacting medium is produced. This system behaves as an almost perfect quantum fluid \cite{Arsene:2004fa,Adcox:2004mh,Back:2004je,Adams:2005dq}. There are also indications from the RHIC experiments that at these energies the initial state of the colliding nuclei may be a color  glass condensate~\cite{Iancu:2000hn,Arsene:2004fa,Adams:2006uz}. The presence of such a state may affect  the spatial distribution of partons within the nucleus, particularly at high pseudorapidity \cite{PhysRevC.74.044905,Hirano:2005xf}. Measuring the distribution of transverse energy over a wide pseudorapidity range sheds light also on the longitudinal expansion of the system. In 2010, the Large Hadron Collider (LHC) accelerated heavy ions to energies 14 times higher than RHIC in order to produce matter at energy densities never achieved before. One of the basic measurements in this new regime is that of the energy distribution of all the produced particles, which is connected to the initial energy and entropy densities of the produced matter. At lower energies, the measured rapidity distributions of particles are generally well described by Gaussians. The widths of these distributions are consistent with the predictions of Landau hydrodynamics, i.e. $\sigma_y = \sqrt{\ln \gamma}$, where $\gamma$ is the Lorentz factor of the colliding beams~\cite{Landau:1953gs,Bearden:2004yx,Arsene:2009jg,Arsene:2009nr,Landau:1953gs,Belenkij:1956cd}. The rapidity variable, $y \equiv \tanh^{-1} (v_{z}/c)$, where $v_{z}$ is the velocity of the particle along the beam direction $z$, provides a way to describe the longitudinal distribution of matter created in these collisions. Calorimeters measure only the energy deposited at various angles and therefore the data are presented in terms of the distribution of energy in pseudorapidity, $\eta$. Pseudorapidity is defined as $\eta \equiv -\ln[\tan(\theta/2)]$, with $\theta$ the polar angle with respect to the $z$ axis. When the momentum of a particle is larger than its mass, its $\eta \approx y$.

The Compact Muon Solenoid (CMS) experiment is a general-purpose detector designed to study hadron collisions at the\TeV scale \cite{JINST}. In particular, it has almost hermetic calorimetry that is sensitive to the distribution of energy over nearly the complete angular range. The transverse energy is defined by $\ET=\sum_{i} E_i \sin\theta_i$, where $E_i$ is the energy seen by the calorimeter for the $i^{th}$ particle, $\theta_i$ is the polar angle of particle $i$, and the sum is over all particles emitted into a fixed solid angle in an event. The quantity $\rd\ET/\rd\eta$ is an approximately Lorentz invariant measure of the energy distribution. The transverse energy is studied as a function of the geometry of the collision, i.e. the centrality, of the heavy-ion interaction. Finally, comparisons are made with lower energy data and theoretical models.

The central feature of the CMS apparatus is a superconducting solenoid, of 6\unit{m} internal diameter, providing a magnetic field of 3.8\unit{T}. Within the central field volume are the silicon pixel and strip trackers, lead-tungstate crystal electromagnetic calorimeter (ECAL) and the brass-scintillator hadron calorimeter (HCAL). These calorimeters are physically divided into the barrel and endcap regions covering together the region of $|\eta|<3.0$. The Hadronic Forward (HF) calorimeters cover $|\eta|$ from 2.9 to 5.2. The HF calorimeters use quartz fibers embedded within a steel absorber.   The CMS tracking system, located inside the calorimeter, consists of pixel and silicon-strip layers covering $|\eta|<2.5$. A set of scintillator tiles, the Beam Scintillator Counters (BSC), are mounted on the inner side of the HF calorimeters to trigger on heavy-ion collisions and reject beam-halo interactions. In addition, two Zero Degree Calorimeters (ZDC) are used for systematic checks. For more details on CMS see~\cite{JINST}.

In 2010, CMS recorded PbPb collision data corresponding to an integrated luminosity of 7.36$\,\mu {\rm b}^{-1}$  of which 0.31$\,\mu {\rm b}^{-1}$ were included in this analysis. This luminosity selection provided a data sample with negligible  statistical uncertainties. Minimum bias inelastic PbPb collisions were selected by requiring that either the HF or BSC counters detected a signal on both sides of the interaction point. In the analysis at least two reconstructed tracks were required to form a vertex within ${\pm}25\unit{cm}$ of the nominal interaction point along the beam line and within a radius of 2\unit{cm} measured perpendicular to the beam relative to the average vertex position. Large-multiplicity beam-background events were removed  by requiring the compatibility of the observed pixel-cluster lengths with the hypothesis of a PbPb interaction at the estimated vertex. Finally, events containing beam-halo muons were eliminated by  a timing requirement on the BSC counters on opposite sides of the interaction point. The total event selection efficiency of the minimum bias trigger for hadronic PbPb interactions was found to be $(97\pm3)$\%~\cite{Chatrchyan:2011sx}.

Events were sorted into different centrality classes. The centrality of heavy-ion interactions is related to the number of participating nucleons and hence to the energy released in the collisions. In CMS, the centrality is defined as percentiles of the energy deposited in the HF. The most central/peripheral event class, i.e. (0--2.5)\%/(70--80)\% in this analysis, has a large/small number of participants and a large/small energy deposit in HF. In order to estimate the mean number of participating nucleons ($\langle N_{\text{part}} \rangle$) and its systematic  uncertainty for each centrality class, a Glauber model of the nuclear collision was used~\cite{glauber,phobos_glauber,Chatrchyan:2011pb}.

The data were corrected for detector acceptance and inefficiencies using correction factors $C(|\eta|)$ estimated from the {\sc hydjet 1.8}~\cite{Lokhtin:2005px} Monte Carlo (MC) event generator coupled to a  \GEANTfour~\cite{Agostinelli:2002hh} CMS detector simulation. These correction factors were calculated as the ratio of MC predictions at the particle level and the detector level for each centrality class. The correction factor $C(|\eta|)\approx$ 1.6 for $|\eta|<$ 2, falls to $\approx$ 1.1 by  $|\eta|$ = 4 and then rises to 2 at $|\eta|\approx5$. The non-linearity of the calorimeter response and the effect of the magnetic field cause the $C(|\eta|)$ to depend upon the $\pt$ spectra,  the ratio of charged and neutral particles, and the mixture of mesons and baryons. The value of $C(|\eta|)$ increases if the assumed spectra shifts to lower $\pt$ or if the ratio of charged to neutral particles is larger. To estimate the systematic uncertainties in $C(|\eta|)$, two tunes of {\sc hydjet} ({\sc 1.6} and {\sc 1.8})  were used. {\sc hydjet 1.8}, $\langle  \pt \rangle_\text{ch} =0.66\GeVc$, was tuned to LHC spectra and particle yields as measured by the ALICE collaboration~\cite{Aamodt:2010cz} and successfully tested against a wide range of RHIC data. {\sc hydjet 1.6}, $\langle \pt \rangle_{\rm ch} =0.57\GeVc$, was only tuned to RHIC data~\cite{Lokhtin:2005px}. At central pseudorapidity the fraction of $\ET$ carried by charged pions, kaons, protons and antiprotons is 0.60 for {\sc hydjet 1.6} and 0.62 for {\sc hydjet 1.8}. The results were cross checked using data taken with no magnetic field and in addition, for B = 3.8\unit{T}, data from tracks with $\pt>900\MeVc$ were combined with energy clusters in the calorimeters to identify different types of particles and measure their energy. Since muons and neutrinos carry a negligible fraction of the total transverse energy and deposit almost no signal in the calorimeters they are not considered in this analysis and no correction factors are applied to account for them. The corrected transverse energy for $N$ analyzed events is obtained as
\begin{equation}
\frac{\rd\ET}{\rd\eta} (|\eta|) = C(|\eta|) \cdot \frac{\sum_j E_{\mathrm{T}, j}  (|\eta|)}{N \times (2\times \Delta \eta)}
\end{equation}
where the sum over $j$ covers all calorimeters cells located within the $\eta$ range $\Delta \eta$ and $E_{T, j}$ is the transverse energy measured in a particular cell $j$. Note that for this summation no threshold is applied to the individual calorimeters cells.
Several sources of systematic uncertainties were studied and their effects are summarized in Table~\ref{tab:SysErrors} and described below. Energy Scale: all of the calorimeters were initially calibrated with test beam data and radioactive sources. For the barrel and inner endcap calorimeters these calibrations were refined using isolated charged hadrons whose momentum was reconstructed in the tracker. For the HF calorimeter the energy scale was cross-checked by reconstructing $\cPZ\to\Pep\Pem$ in $\Pp\Pp$ collisions where either the positron or the electron was recorded in the ECAL and tracker. Symmetry about $\eta$: for PbPb collisions the corrected $\rd\ET/\rd\eta$ should  be symmetric about $\eta=0$. The values of $\rd\ET/\rd\eta$ for positive and negative $\eta$ were found to differ by at most 0.5\%. This close agreement implies that one can use the average of the two results as a best estimate of $\rd\ET/\rd\eta$. Vertex distribution: the $z$ distribution of the vertices is Gaussian with $\sigma_{z}\approx6.1\unit{cm}$. To test the sensitivity of $\ET$ to the position of the interaction vertex along the beam line, $z$,  the dataset was divided  into two samples with $|z|< 10\unit{cm}$ and $10\unit{cm}<|z|<25\unit{cm}$, respectively. The $\ET$ distributions of the two samples differ by less than 2\%. Auto-correlations: since HF is used both to calculate  centrality and to measure $\ET$ for each centrality class, there is an auto-correlation in the measurement. This effect was estimated to be less than 1.5\% by using a combination of the ZDC and pixel detectors to measure centrality. Calorimeter noise: the \GEANTfour simulation of the calorimeters included electronic noise. This noise was measured by studying a sample of events where the trigger only required the presence of clockwise and anticlockwise bunches of lead ions simultaneously in  CMS. The simulation of the noise was checked by comparing the data to the simulated signal from a {\sc geant4} simulation of the most peripheral events in the data set. Any discrepancy in the simulation of the noise corresponds to a corrected average $\ET$ per event of less than 5.8\GeV for   $|\eta| \le 2.65$ and 1.2\GeV for $2.65 < |\eta| \le 5.2$. This is only significant compared to the signal for $\langle N_{\text{part}} \rangle \le 30$. HF MC description: takes into account different ways of describing the dead areas of the HF detector. Centrality determination: the systematic uncertainty related to the centrality determination is only applied to the results that are normalized by $\langle N_{\text{part}} \rangle$.

\begin{table}[htb]
\centering
\topcaption{Systematic uncertainties for two $|\eta|$ regions in \% for the most central (0--2.5)\% ($\langle N_{\text{part}} \rangle$~=~394) and most peripheral (70--80)\% ($\langle N_{\text{part}} \rangle=16$) collisions.}
\begin{tabular}{|c|c|c|c|c|}
\hline
&\multicolumn{2}{c|}{$|\eta| \le 2.65$}&\multicolumn{2}{c|}{$2.65 < |\eta| \le 5.2$}\\
$\langle N_{\text{part}} \rangle$ &16& 394&16& 394\\
 \hline
\hline
Energy scale             &   2          &2       & 10 &10  \\
MC model                  &   1.2--12   & 1.2--4.9     &1--6.8 & 1--2.3  \\
Vertex               &   2           &2           &2 & 2  \\
$\eta$ symmetry   &   0.5        &0.5        & 0.3 &0.3   \\
Auto-correl.                 &   1.5        &1.5        & 1.0 &1.0   \\
Calo. noise                  &  14--18    &0.3        &4--7.3 & 0.1--0.2  \\
HF MC & --&--&1.5--9&1.5--9\\
Centrality                  & 6.7        &0.5   & 6.7  &0.5 \\
\hline
Total                   &   14--22   & 3.5--5.9  & 11--17 &10--14   \\
\hline
\end{tabular}
\label{tab:SysErrors}

\end{table}

\begin{figure}[h]
\centering
\includegraphics[width=\cmsFigWidth]{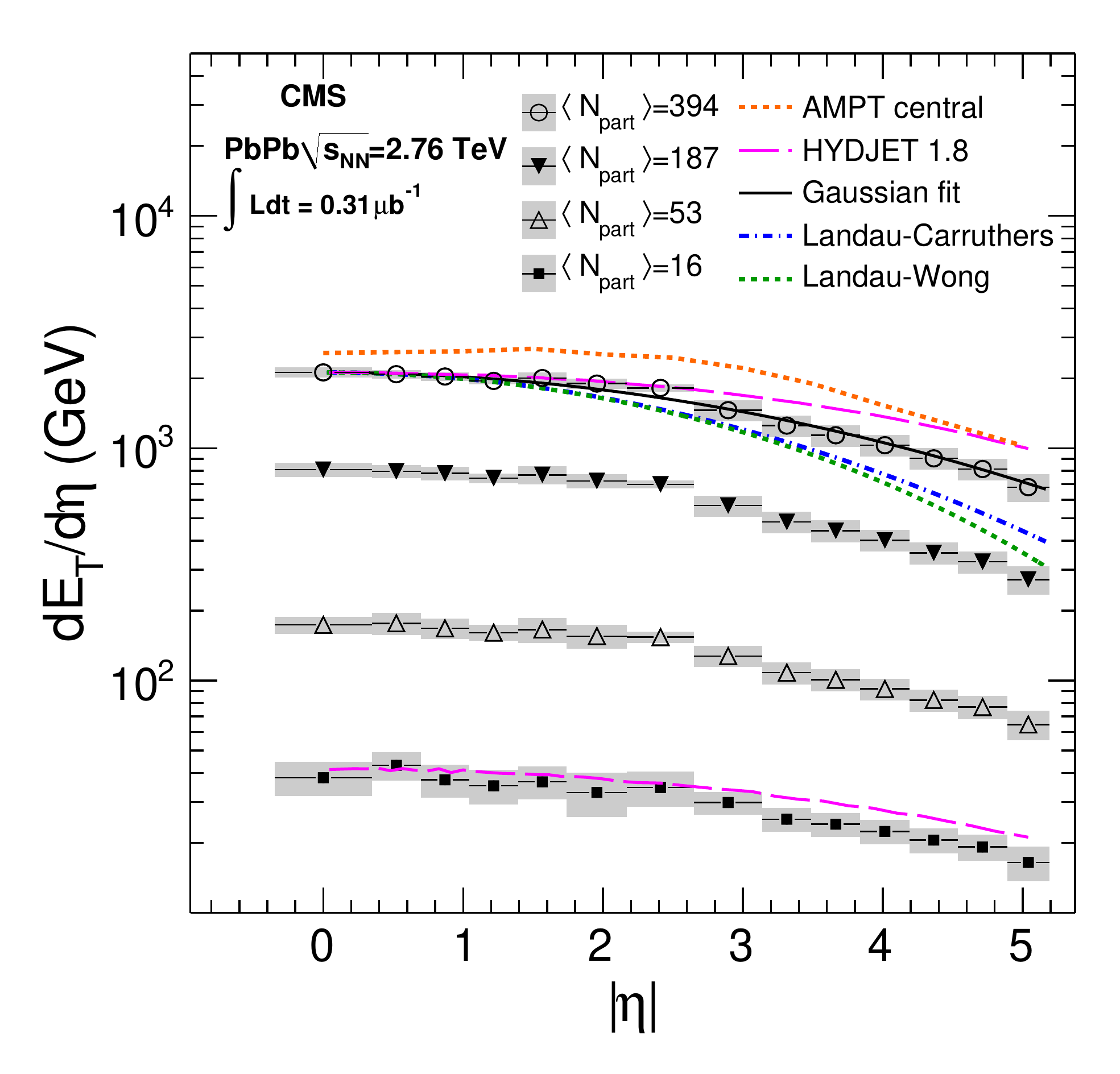}
   \caption{\label{dEtdEta} Transverse energy density versus $|\eta|$ distribution for a range of centralities of (0--2.5)\%, (20--30)\%, (50--60)\% and (70--80)\%. The boxes show the total systematic uncertainties. The statistical uncertainties are negligible. Also shown are a Gaussian fit and the predictions of various models (see text). The AMPT events are for perfectly central collisions.
    }
\end{figure}

Figure~\ref{dEtdEta} shows the $|\eta|$ dependence of the transverse energy density for four selected ranges of centrality. For the most central collisions ($\langle N_{\text{part}} \rangle=394$) $\rd\ET/\rd\eta$ reaches 2.1\TeV at $\eta=0$. This is much larger than the value of 0.61\TeV measured at $\sqrt{s_{_\mathrm{NN}}} = 200$\GeV~\cite{Adler:2004zn}. At lower center-of-mass energies the pion multiplicity distributions are reasonably well described by Gaussians in rapidity with widths that are consistent with Landau-Carruthers hydrodynamics~\cite{Carruthers:1973rw,Carruthers:1972}.  Since the mean $\pt$ of all particle species depends only weakly on rapidity, this implies that $\rd\ET/\rd y$ is roughly Gaussian in rapidity at $\sqrt{s_{_\mathrm{NN}}}$  = 200\GeV.  Recently, Wong has improved the formulation of Landau hydrodynamics~\cite{Wong:2008ex}. This new formulation gives a better description of the 200\GeV RHIC data. At $\sqrt{s_{_\mathrm{NN}}}$  = 2.76\TeV and for $|\eta|<5.2$ the $\rd\ET/\rd\eta$ is consistent with a Gaussian (black solid line) with $\sigma_\eta = 3.4 \pm 0.1$ for the most central collisions. The Gaussian and Landau curves in Fig.~\ref{dEtdEta} are normalized to the CMS data at $\eta=0$. Both the Landau-Carruthers (blue dashed),  and Landau-Wong (green dotted) formulations have distributions that are narrower than the data. Therefore the longitudinal expansion of the system is stronger than that predicted from either model. {\sc hydjet 1.8}, shown by the purple dashed line,  has been tuned to LHC data in the small $|\eta|$ region. It gives a good description of  $\rd\ET/\rd\eta$ at small $|\eta|$ but overestimates the data at large $|\eta|$ for central collisions. The {\sc ampt} (A Multi Phase Transport) model~\cite{Zhang:1999bd,Lin:2004en} (orange dashed line) overestimates $\rd\ET/\rd\eta$ for central collisions but is in rough agreement with the shape of $\rd\ET/\rd\eta$. Integrating $(\rd\ET/\rd\eta)/(\langle N_{\text{part}} \rangle/2)$ over $\eta$ between $-5.2$ and 5.2  gives a total measured $\ET$ per participant pair of $80 \pm 4$\GeV for the most central events. This serves as a lower limit for the total transverse energy per nucleon pair. Extrapolating to the full phase space gives a total transverse energy per pair of participating nucleons of $91\pm 5$\GeV for the most central events.  It is clear from Fig.~\ref{dEtdEta} that the magnitude of $\rd\ET/\rd\eta$ increases rapidly with the number of nucleons participating in the collision.\\

\begin{figure}[htp]
 \centering
    \includegraphics[width=\cmsFigWidth]{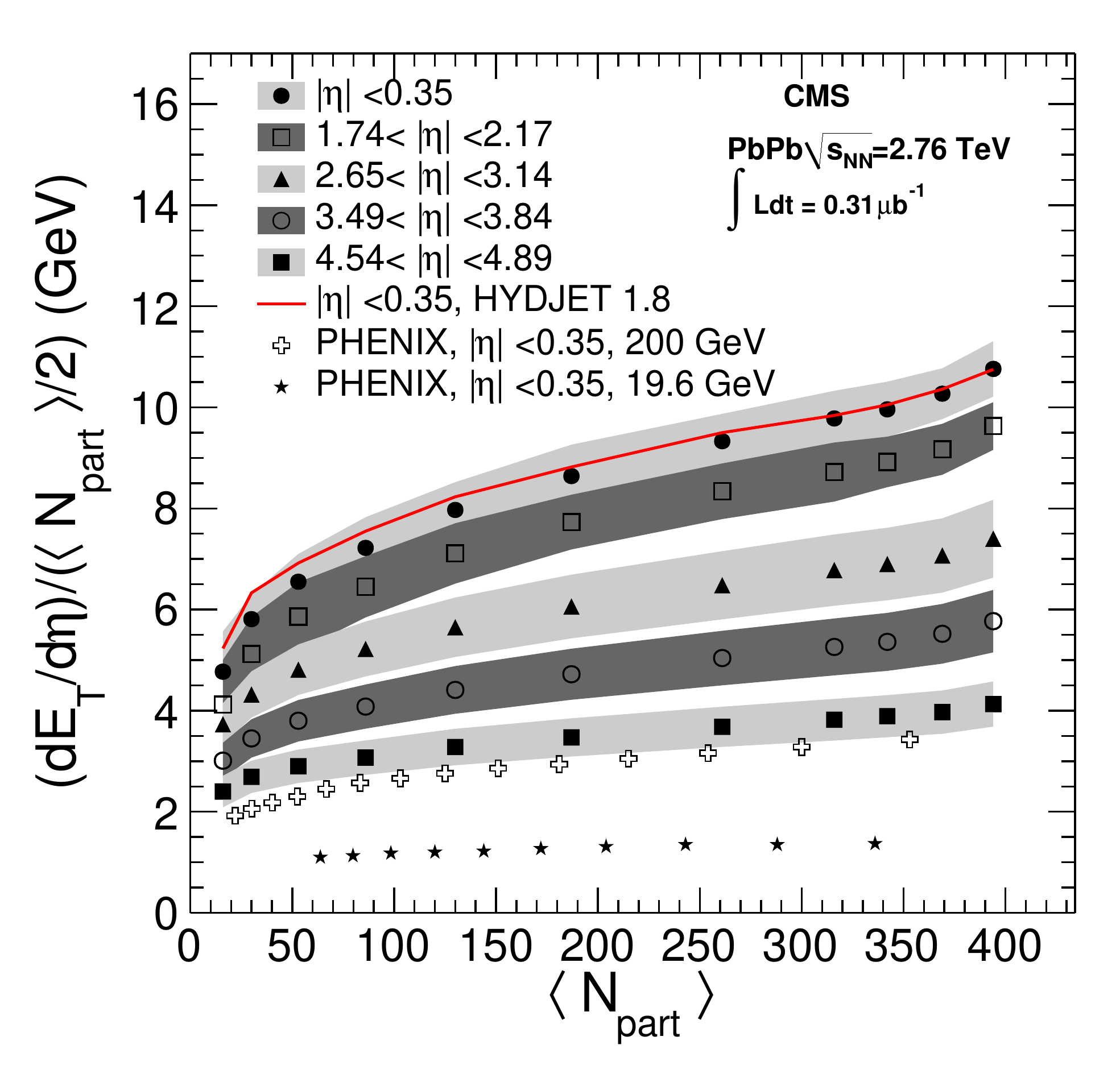}
   \caption{\label{EtvNpart} Transverse energy density normalized by ($\langle N_{\text{part}} \rangle/2$) versus $\langle N_{\text{part}} \rangle$
   for  PbPb collisions at $\sqrt{s_{_\mathrm{NN}}} = 2.76\TeV$ at several values of $|\eta|$.
   The bands show the total systematic uncertainties.  The statistical uncertainties are negligible. For $\eta=0$ lower-energy PHENIX data and results from the {\sc hydjet 1.8} model are also shown. }
\end{figure}

Figure~\ref{EtvNpart} shows the evolution of $(\rd\ET/\rd\eta)/(\langle N_{\text{part}} \rangle/2)$ with
$\langle N_{\text{part}} \rangle$ for several $|\eta|$ regions. At all $|\eta|$ values $(\rd\ET/\rd\eta)/(\langle N_{\text{part}} \rangle/2)$ increases with $\langle N_{\text{part}} \rangle$. This figure shows that the $\langle N_{\text{part}} \rangle$ dependence of transverse energy density changes as a function of pseudorapidity. This effect can be quantified by comparing peripheral (60--70)\% ($\langle N_{\text{part}} \rangle=30$)  to central (0--2.5)\% collisions ($\langle N_{\text{part}} \rangle=394$) at various pseudorapidities.  The ratio of peripheral to central $(\rd\ET/\rd\eta)/(\langle N_{\text{part}} \rangle/2)$ changes from $54 \pm 2 \%$ at $\eta = 0$ to $68 \pm 2\%$ at $|\eta| =  5.0$. The PHENIX collaboration at RHIC has studied transverse energy density in AuAu collisions for $|\eta|<0.35$ over a wide range of centralities and for $\sqrt{s_{_\mathrm{NN}}}$ from 19.6\GeV to 200\GeV \cite{Adler:2004zn}. At $\sqrt{s_{_\mathrm{NN}}}=19.6\GeV$ $(\rd\ET/\rd\eta)/(\langle N_{\text{part}} \rangle/2)$ at $\eta=0$ increases by a factor of $1.25\pm0.17$ as  $\langle N_{\text{part}} \rangle$ increases from 63.8 to 336. At $\sqrt{s_{_\mathrm{NN}}}=2.76\TeV$ this factor is found to be $1.47\pm0.13$ for a similar range of $\langle N_{\text{part}} \rangle$. At $\sqrt{s_{_\mathrm{NN}}}=2.76\TeV$, the {\sc hydjet 1.8} code gives a good description of the centrality dependence of $\rd\ET/\rd\eta$ at $\eta=0$.

\begin{figure}[Hhtb]
\centering
   \includegraphics[width=\cmsFigWidth]{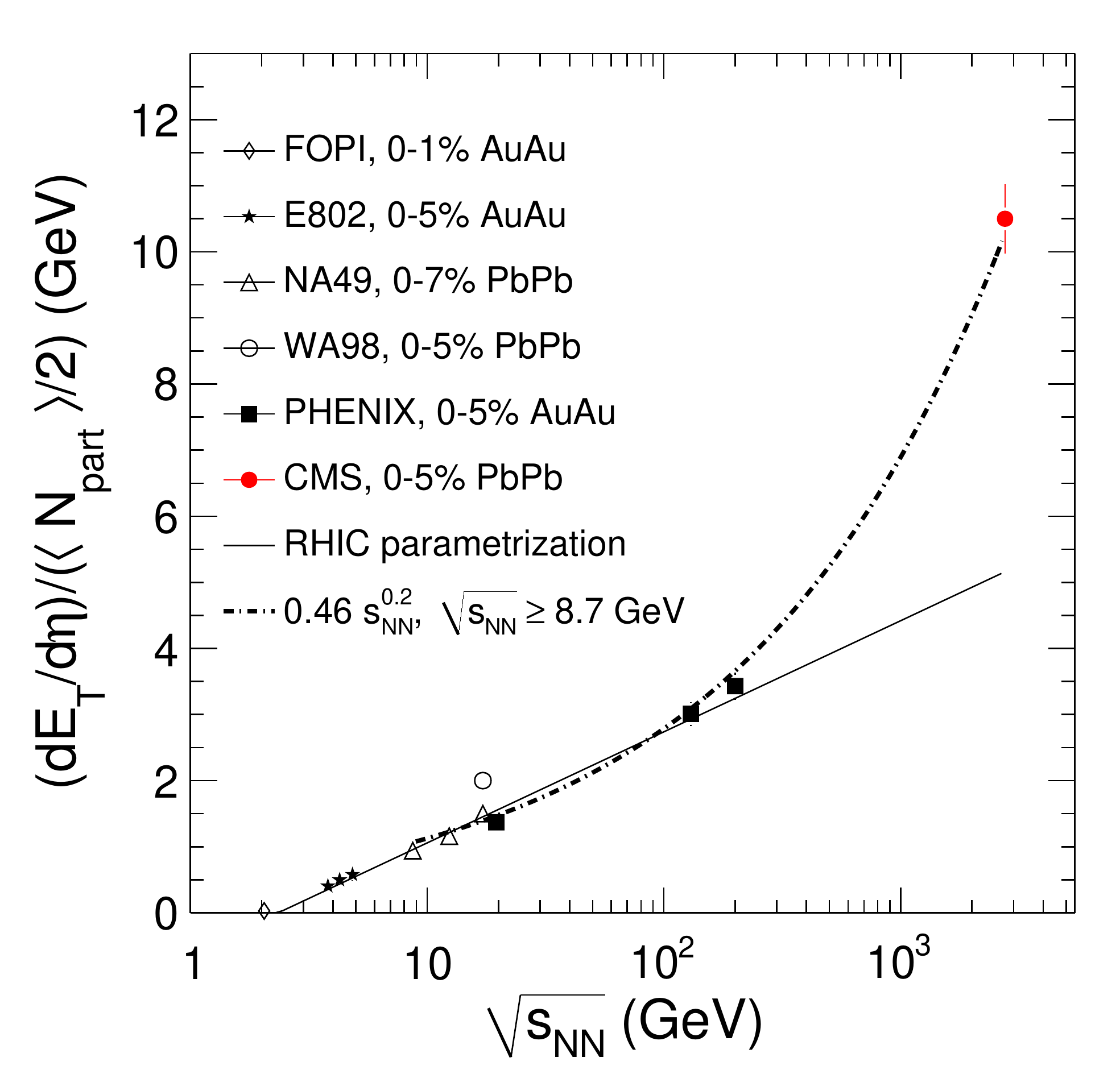}
   \caption{Transverse energy density normalized by ($\langle N_{\text{part}} \rangle/2$) for central collisions at $\eta=0$ versus $\sqrt{s_{_\mathrm{NN}}}$. The vertical bar on the CMS point is the full systematic uncertainty. The statistical uncertainty is negligible. The solid line is a parametrization of  lower-energy data compiled by the PHENIX Collaboration, see
   \cite{Adler:2004zn,Abbott:2001gd,Aggarwal:2000bc,Bachler:1999hu,Afanasiev:2002fk,Afanasiev:2002mx,Reisdorf:1900zza,Pelte:1997rg,Hong:2001tm}. The dashed line corresponds to a power law fit $s_{_\mathrm{NN}}^{n}$ for $\sqrt{s_{_\mathrm{NN}}} \geq8.7\GeV$.}
   \label{EtvsRootS}
\end{figure}

Figure~\ref{EtvsRootS}  shows the energy dependence of $(\rd\ET/\rd\eta)/(\langle N_{\text{part}} \rangle/2)$ for central collisions at $\eta=0$. For the top 5\% most central events $(\rd\ET/\rd\eta)/(\langle N_{\text{part}} \rangle/2)$ reaches $10.5 \pm 0.5$\GeV at $\sqrt{s_{_\mathrm{NN}}}$  = 2.76\TeV. The $\ET$ rises more quickly with the center-of-mass energy than the logarithmic dependence used to describe data up to $\sqrt{s_{_\mathrm{NN}}} = 200$\GeV~\cite{Adler:2004zn}. For energies between 8.7\GeV and 2.76\TeV, $\rd\ET/\rd\eta$ at  $\eta=0$ can be reproduced by a power-law dependence of the type $s_{_\mathrm{NN}}^{n}$ with $n \approx0.2$. A similar effect has been seen in the measurement of the $\sqrt{s_{_\mathrm{NN}}}$ evolution of the charged particle multiplicity~\cite{ALICE_centraldNdeta,Chatrchyan:2011pb}. The $(\rd\ET/\rd\eta)/(\langle N_{\text{part}} \rangle/2)$ increases by a factor of $ 3.07\pm0.24$ from $\sqrt{s_{_\mathrm{NN}}}$  = 200\GeV to 2.76\TeV. This is to be compared to a factor of $2.17\pm0.15$ for the pseudorapidity density, $(\rd N_\text{ ch}/\rd\eta)/(\langle N_{\text{part}} \rangle/2)$~\cite{Aamodt:2010cz,Chatrchyan:2011pb,Adler:2004zn}. For the 5\% most central collisions CMS has measured $\rd\ET/\rd\eta$ = $2007 \pm 100$\GeV and $\rd N_\text{ch}/\rd\eta$ = $1612 \pm 55$~\cite{Chatrchyan:2011pb}. Dividing the measured transverse energy by the observed charged particle multiplicity for the same centrality gives a transverse energy per charged particle of $1.25\pm0.08$\GeV at $\sqrt{s_{_\mathrm{NN}}}$  = 2.76\TeV. This compares to  0.88 $\pm$ 0.07\GeV at $\sqrt{s_{_\mathrm{NN}}}$  = 200\GeV~\cite{Adler:2004zn}.

The sum of the transverse energies of all particles produced in the event depends upon both the entropy and temperature of the system. Using geometrical considerations, Bjorken~\cite{Bjorken} suggested that the energy density per unit volume in nuclear collisions could be estimated from the energy density per unit rapidity. A commonly used estimate of energy density is given by~\cite{Adler:2004zn}
\begin{equation}
\epsilon = \frac{1}{A c  \tau_0}  J(y,\eta) \frac{\rd\ET}{\rd\eta}.
\label{eq:EnergyDensity}
\end{equation}
where $A$ is the overlap area of the two nuclei and $\tau_0$ is the formation time of the produced system.  The Jacobian $J(y,\eta)$ depends on the momentum distributions of the produced particles. In the limit that the rest mass of the particles are much smaller than their momenta $J(y,\eta)=1$. The average Jacobian was calculated using {\sc hydjet 1.8} for $|\eta| <0.35$. For central collisions at $\sqrt{s_{_\mathrm{NN}}} = 2.76$\TeV, $J(y,\eta)$ = 1.09. This compares to 1.25 at $\sqrt{s_{_\mathrm{NN}}} = 200$\GeV \cite{Adler:2004zn}. For the top 5\% most central collisions this formula gives $ \epsilon=14\GeV/\mathrm{fm}^3$ at a time $\tau_0=1\unit{fm}/c$ and for a transverse surface of $A = \pi \times (7\unit{fm})^2$~\cite{Adler:2004zn}. This is a factor of 2.6 times larger than the energy density calculated at $\sqrt{s_{_\mathrm{NN}}} = 200$\GeV~\cite{Adler:2004zn}.

In summary, for the most central PbPb collisions at $\sqrt{s_{_\mathrm{NN}}} = 2.76$\TeV, the maximum of the transverse energy distribution has been found to be  2.1\TeV at $\eta=0$. Even at a very forward pseudorapidity of $|\eta|=5.0$, $\rd\ET/\rd\eta$ and hence the energy density of the produced system at the LHC is larger than that measured for $\eta=0$ at RHIC. At 2.76\TeV, the shape of $\rd\ET/\rd\eta$ is consistent with a Gaussian function of width $\sigma_\eta = 3.4 \pm 0.1$ for central collisions. This distribution is wider than the prediction of Landau hydrodynamics but narrower than that given by the {\sc hydjet 1.8} simulation. The  $(\rd\ET/\rd\eta)/(\langle N_{\text{part}} \rangle/2)$ increases with $\langle N_{\text{part}} \rangle$ at all pseudorapidities. The ratio of transverse energy in peripheral compared to central collisions increases by a factor of 1.26 $\pm$ 0.06 from $\eta=0$ to $|\eta|=5$. The transverse energy density at $\eta=0$ grows  more rapidly with the center-of-mass energy than the  logarithmic scaling with $\sqrt{s_{_\mathrm{NN}}}$ that describes lower-energy data. It also grows faster with energy than the multiplicity, implying a significant increase of the mean transverse energy per particle compared to lower energy data.

\section*{Acknowledgments}

We wish to congratulate our colleagues in the CERN accelerator departments for the excellent performance of the LHC machine. We thank the technical and administrative staff at CERN and other CMS institutes, and acknowledge support from: FMSR (Austria); FNRS and FWO (Belgium); CNPq, CAPES, FAPERJ, and FAPESP (Brazil); MES (Bulgaria); CERN; CAS, MoST, and NSFC (China); COLCIENCIAS (Colombia); MSES (Croatia); RPF (Cyprus); Academy of Sciences and NICPB (Estonia); Academy of Finland, MEC, and HIP (Finland); CEA and CNRS/IN2P3 (France); BMBF, DFG, and HGF (Germany); GSRT (Greece); OTKA and NKTH (Hungary); DAE and DST (India); IPM (Iran); SFI (Ireland); INFN (Italy); NRF and WCU (Korea); LAS (Lithuania); CINVESTAV, CONACYT, SEP, and UASLP-FAI (Mexico); MSI (New Zealand); PAEC (Pakistan); SCSR (Poland); FCT (Portugal); JINR (Armenia, Belarus, Georgia, Ukraine, Uzbekistan); MST and MAE (Russia); MSTD (Serbia); MICINN and CPAN (Spain); Swiss Funding Agencies (Switzerland); NSC (Taipei); TUBITAK and TAEK (Turkey); STFC (United Kingdom); DOE and NSF (USA). Individuals have received support from the Marie-Curie programme and the European Research Council (European Union); the Leventis Foundation; the A. P. Sloan Foundation; the Alexander von Humboldt Foundation; the Associazione per lo Sviluppo Scientifico e Tecnologico del Piemonte (Italy); the Belgian Federal Science Policy Office; the Fonds pour la Formation \`a la Recherche dans l'Industrie et dans l'Agriculture (FRIA-Belgium); the Agentschap voor Innovatie door Wetenschap en Technologie (IWT-Belgium); and the Council of Science and Industrial Research, India.

\bibliography{auto_generated}

\cleardoublepage \appendix\section{The CMS Collaboration \label{app:collab}}\begin{sloppypar}\hyphenpenalty=5000\widowpenalty=500\clubpenalty=5000\textbf{Yerevan Physics Institute,  Yerevan,  Armenia}\\*[0pt]
S.~Chatrchyan, V.~Khachatryan, A.M.~Sirunyan, A.~Tumasyan
\vskip\cmsinstskip
\textbf{Institut f\"{u}r Hochenergiephysik der OeAW,  Wien,  Austria}\\*[0pt]
W.~Adam, T.~Bergauer, M.~Dragicevic, J.~Er\"{o}, C.~Fabjan\cmsAuthorMark{1}, M.~Friedl, R.~Fr\"{u}hwirth\cmsAuthorMark{1}, V.M.~Ghete, J.~Hammer, N.~H\"{o}rmann, J.~Hrubec, M.~Jeitler\cmsAuthorMark{1}, W.~Kiesenhofer, V.~Kn\"{u}nz, M.~Krammer\cmsAuthorMark{1}, D.~Liko, I.~Mikulec, M.~Pernicka$^{\textrm{\dag}}$, B.~Rahbaran, C.~Rohringer, H.~Rohringer, R.~Sch\"{o}fbeck, J.~Strauss, A.~Taurok, P.~Wagner, W.~Waltenberger, G.~Walzel, E.~Widl, C.-E.~Wulz\cmsAuthorMark{1}
% ...
\vskip\cmsinstskip
\dag:~Deceased\\
1:~~Also at Vienna University of Technology, Vienna, Austria\\
% ...

\end{sloppypar}
\end{document}